\documentclass[prl,aps,twocolumn,nofootinbib]{revtex4-1}
\usepackage{amssymb}
\usepackage{amsmath}
\usepackage{amstext}
\usepackage{amsfonts}
\usepackage{bbold}
\usepackage{slashed}
\usepackage{graphicx}
\usepackage[letterpaper, left=2cm, right=2cm, top=2.0cm, bottom=1.8cm]{geometry}

\def\l{\left}
\def\r{\right}
\def\fr{\frac}
\def\la{\label}
\def\d{\partial}

\def\be{\begin{eqnarray}}
\def\ee{\end{eqnarray}}

\begin{document}

\title{An effective field theory approach \\ to tidal dynamics of spinning astrophysical systems} 

\author{Solomon Endlich}
\affiliation{Stanford Institute for Theoretical Physics, \\ Stanford University,
Stanford, CA 94306, USA}
\author{Riccardo Penco}
\affiliation{Physics Department, Center for Theoretical Physics\\
and Institute for Strings, Cosmology, and Astroparticle Physics,\\
  Columbia University, New York, NY 10027, USA}

\begin{abstract}
We develop a description of tidal effects in astrophysical systems using effective field theory techniques. While our approach is equally capable of describing objects in the Newtonian regime (e.g. moons, rocky planets, main sequence stars, etc.) as well as relativistic objects (e.g. neutron stars and black holes), in this paper we focus special attention on the Newtonian regime. In this limit, we recover the dynamical equations for the ``weak friction model'' with additional corrections due to tidal and rotational~deformations.
\end{abstract}

\maketitle

The study of the tidal effects is nearly as old as Newton's theory of gravitation itself. Beginning with the seminal work of G.~ Darwin~\cite{Darwin}, tidal phenomena have been traditionally regarded as a complicated viscoelastic/fluid dynamics problem, and have since been tackled with an ever increasing level of sophistication and ingenuity. Perhaps the most familiar manifestation of tides is the periodic deformation experienced by the Earth as a result of the gravitational interaction with the Moon and the Sun (the rest of the objects in our solar system have a comparatively negligible effect). Using phenomenological models whose parameters have been fit against detailed geophysical data scientists can describe such deformations and, as a result, compile accurate tide tables for locations all over the world.

Over much longer time scales, tides have also a crucial impact on the orbital dynamics, leading to well known---and easily observed---dissipative phenomena such as tidal locking and orbital circularization~\cite{murray1999solar}. For non-terrestrial astrophysical objects these dynamics can provide an indirect probe of the otherwise extremely difficult to measure internal structure of the body.

In this letter we will focus our attention on orbital tidal effects for spinning objects, and we will discuss an alternative approach to this problem based on effective field theory (EFT) techniques. The main idea behind any EFT is that, in a first approximation, physics at large distances should admit a description that is independent of the physics that takes place at much smaller length scales~\cite{Goldberger:2007hy}. In our case, this translates into the statement that the dynamics of astrophysical spinning objects separated by a distance $r$ much larger than their own size $\ell$ can be approximated at a first pass by treating these objects as point-like and neglecting their internal structure. For a given accuracy, this may not be adequate. The power of the EFT framework is that it makes it possible to still work with the ``point-like'' degrees of freedom and yet go beyond this approximation and systematically include finite-size corrections to the dynamics up to any desired order in~$\ell /r$~\cite{Goldberger:2004jt}.

For compact objects of comparable mass (i.e. objects whose typical size is comparable to the Schwarzschild radius $\ell \sim GM/c^2$) in near Keplerian orbits, the ratio $\ell / r$ is also approximately equal to $(v/c)^2$. Consequently, for these objects the expansion in $\ell/r$ is equivalent to a post-Newtonian expansion. This fact forms the basis for the Non-Relativistic General Relativity (NRGR) approach to calculate post-Newtonian corrections to the motion of binary inspirals and the resulting gravitational wave emission (see~\cite{Rothstein:2014sra} and references therein). This formalism was extended in~\cite{Porto:2005ac} to account for the effects of spin, but still assuming that the latter scales as some power of $v /c$. \footnote{Spin in general relativity has a long history; for more information see, for instance, \cite{Obukhov:2012up} and the references contained therein. See also \cite{Binnington:2009bb, Damour:2009vw, Kol:2011vg, Bini:2012gu} for recent developments in finite size effects in compact objects, and \cite{Levi:2015msa} for computations of spin dynamics in NRGR to higher post-Newtonian order.  Notice however, that these papers are concerned exclusively with \emph{conservative} aspects of spin dynamics. In contrast, we will focus on \emph{dissipative} phenomena.} 

This letter represents a first step beyond NRGR, in that we consider non-compact objects for which $\ell /r, v/c$ and the spin are in principle independent expansion parameters.  To be more precise, rather than the spin defined in~\cite{Porto:2005ac}, we will find it convenient to use the ratio $\Omega / \Omega_0$ between the angular velocity and the typical (normal) frequency of the object as our third expansion parameter. Our starting point will be the action for a relativistic spinning object coupled to gravity derived in~\cite{Delacretaz:2014oxa}. In principle, this action contains all possible terms that are compatible with the symmetries (which include both space-time symmetries as well as symmetries of the object itself). In practice, only a finite number of terms are relevant at any given order in $\ell /r$, $v/c$ and $\Omega / \Omega_0$, and thus only a finite number of coefficients---so called ``Wilson coefficients''---are needed to describe the orbital dynamics of a particular body to some specified accuracy over a set time period. The equations of motion in the absence of dissipation can then be easily obtained by varying the action.

In order to capture dissipative effects such as the ones associated with tides, we will need to include additional degrees of freedom in the action~\cite{Goldberger:2005cd,Porto:2007qi,Endlich:2012vt}. This is because dissipation can be thought of as a process in which energy is transferred from the ``large distance'' degrees of freedom (angular and linear position in our case) to the ``microscopic'' ones. Dissipative corrections to the equations of motion can be obtained by averaging over such microscopic degrees of freedom. Within the EFT framework, this averaging procedure can be made precise and carried out in complete generality, without resorting to a concrete model for the microscopic physics. 
\\

\noindent \textsc{Conventions:} we will work in the vierbein formalism with ``mostly plus'' metric signature. We will denote space-time indices with $\mu,\nu,\lambda ...$, Lorentz indices with $a,b,c,...$, and spatial components of the latter with $i,j,k, ...$ . Finally, a boost tensor is parametrized as follows: $B^0{}_0 = \gamma, B^0{}_i = B^i{}_0 = \gamma \beta^i, B^i{}_j = \delta^i_j + (\gamma - 1) \hat \beta^i \hat \beta_j$, where as usual  $\gamma = (1 - \beta^2)^{-1/2}$. Where appropriate, tildes indicate quantities in the comoving frame while no tilde indicates that the quantity is in the lab frame. 
\\


\noindent \textsc{Effective Action.}
In order to describe the motion of a spinning object in a Lorentz invariant fashion, we need to introduce a tensor $\Lambda^a{}_b$ describing a local Lorentz transformation that takes a comoving frame (i.e. a frame embedded in the rigid body) and turns it into an inertial frame (i.e. the lab frame). We can parametrize such Lorentz transformation as the product of a boost and a rotation, $\Lambda^a{}_b = B^a{}_c (\beta^i) R^c{}_b (\theta^i)$ where the $\theta$'s are Euler angles describing the orientation of the rigid body, and
\be
\label{beta_condition}
\beta^i = \fr{\d_\tau x^\mu e_\mu{}^i}{\d_\tau x^\nu e_\nu{}^0}
\ee
with $\tau$ the proper time and $e_\mu{}^a$ the vierbein (which as usual is related to the metric by $g_{\mu\nu} = e_\mu{}^a e_\nu{}^b \eta_{ab}$). Using the vierbein, one can define the spin connection as usual:
\be
\omega_\mu{}^{ab} = e^{\nu a} \d_{[\mu} e_{\nu]}{}^b + \tfrac{1}{2} e_{\mu c} e^{\nu a}e^{\lambda b}\d_\lambda e_\nu{}^c - (a \leftrightarrow b).
\ee
Our action for a spinning object coupled to gravity is then given by
\be \la{action}
S = \int d \tau L (\tilde \Omega_i, \tilde C_{abcd}, \tilde{\nabla}_a, \tilde{\mathcal{O}}),
\ee
where $\tilde \Omega_i$ are the components of (a relativistic generalization of) the angular velocity in a comoving frame, which are defined as
\be
\label{omega_def}
\tilde \Omega_i = - \tfrac{1}{2} \epsilon_{ijk} \Lambda_a{}^j (\eta^{ab} \d_\tau + \d_\tau x^\mu \omega_\mu{}^{ab}) \Lambda_b{}^k ,
\ee
the $\tilde C_{abcd}$'s  are the components of the Weyl tensor\footnote{Neglecting self energies, the leading order equations of motion for the gravitational field are simply $R_{\mu \nu}=0$. Hence, any term proportional to $R_{\mu\nu}$ can be removed by a field redefinition of the metric, and therefore we only need the traceless components of the Riemann tensor---i.~e.~ the Weyl tensor~\cite{Goldberger:2004jt, Damour:1998jk}.} in the same comoving frame, i.e.
\be
\tilde C_{abcd} = (\Lambda^{-1})_a{}^{a'}(\Lambda^{-1})_b{}^{b'}(\Lambda^{-1})_c{}^{c'}(\Lambda^{-1})_d{}^{d'} C_{abcd},
\ee
and $\tilde \nabla_a = (\Lambda^{-1})_a{}^b \nabla_b$ are covariant derivatives evaluated in the comoving body frame. Finally, $\tilde{\mathcal{O}}$ stands for an infinite collection of composite operators that represent all the additional degrees of freedom of the object which are neither its position nor its orientation. By including in the action all possible operators allowed by the symmetries, we will be able to account for such degrees of freedom in a model-independent way. 

Despite appearances, the action (\ref{action}) is completely Lorentz invariant. The reason why this is not manifest is because Lorentz symmetry is realized non-linearly. This action was originally obtained in~\cite{Delacretaz:2014oxa}, and we refer the reader to that paper for further details on its derivation and symmetry properties.

In the case of an object that is spherically symmetric at rest, only rotationally invariant operators are allowed in the action (\ref{action}). This is because, as we will discuss later on, any departure from sphericity caused by rotation is associated with higher derivative terms in the action rather than with terms that explicitly violate rotational invariance. Moreover, higher derivative terms are associated with higher order correction in the $\ell / r, v /c$ and $\Omega / \Omega_0$ expansions.\footnote{The scales $\ell$ and $\Omega_0$ do not appear explicitly in our action, but they determine the characteristic size of the dimensionful coupling constants.} Thus, if we are only interested in the first few terms in this expansion, the most relevant terms in the action are the following:
\be \la{SO(3) action}
S &=& \int d \tau \bigg\{ - m c^2 + \fr{I}{2} \tilde \Omega_i \tilde \Omega^i + \fr{J}{4} (\tilde \Omega_i  \tilde \Omega^i)^2 + \cdots \nonumber \\
&& + n_\Omega \tilde \Omega^i \tilde \Omega^j \tilde C_{0i0j} + n_E \tilde C_{0i0j} \tilde C^{0i0j} + n_B \tilde C_{0ijk} \tilde C^{0ijk} + \cdots \nonumber   \\
&& + \tilde{\mathcal{O}}_E^{ij} \tilde C_{0i0j} + \tilde{\mathcal{O}}_B^{ijk}  \tilde C_{0ijk} + \cdots \bigg\}. 
\ee
The first line contains all the terms necessary to describe the dynamics of a relativistic spinning object even in the absence of gravity; the second line contains terms known as \emph{finite-size terms}~\cite{Goldberger:2004jt}, which encode the fact that the object is not truly point-like; finally, the last line contains the dissipative terms that couple the degrees of freedom we are interested in (gravity, plus linear and angular position of the object) to those that we are not keeping track of explicitly and are ultimately responsible for dissipation. Notice that such interactions should vanish on a flat space-time, since tidal dissipation occurs only in the presence of gravitational interactions. Following~\cite{Goldberger:2005cd, Damour:2009vw}, we have explicitly separated the Weyl tensor into its ``electric'' and ``magnetic'' parity components. Consequently, the coefficients $n_E$ and $n_B$ can be understood as the gravitational electric and magnetic susceptibility for the object. $n_\Omega$ has a similar interpretation which we will comment on in the next section.
In principle, the full action will also contain a part that specifies the dynamics of the degrees of freedom associated with the $\tilde{\mathcal{O}}$'s. We have omitted this part for notational simplicity because its detailed form is unknown and, at the same time, irrelevant for what follows. 

As mentioned in the introduction, the action (\ref{SO(3) action}) contains in principle an infinite number of terms. What makes such a theory  predictive is the fact that only a finite number of them contribute at any given order in $\ell / r, v /c$ and $\Omega / \Omega_0$. For instance, terms with higher powers of $\tilde \Omega^i$ in the first line yield higher order corrections in $\Omega / \Omega_0$. When $\theta^i =$ constant, the action (\ref{SO(3) action}) reduces to the action of NRGR~\cite{Goldberger:2004jt} in the presence of dissipation~\cite{Goldberger:2005cd}. In this limit for compact objects, the systematics of the $v/c$ expansion has been fully worked out~\cite{Goldberger:2007hy}. The richer perturbative structure that become necessary when $\ell / r$ is different from $(v /c)^2$ will be more fully discussed elsewhere~\cite{futureus}. Finally, it is important to stress that our formalism is different from NRGR with spin~\cite{Porto:2005ac} because our action contains only physical degrees of freedom and as such there is no need to impose any ``spin supplementary condition''~\cite{Hanson:1974qy}. For interested readers familiar with the more standard relativistic spin formalism we clarify this last point with further discussion in the Appendix. 
\\




\noindent \textsc{Newtonian Limit.} In the rest of this letter we will assume that $v/c$ is much smaller than $\ell /r$ and $\Omega/\Omega_0$, as is the case for typical non-compact objects. We will therefore set $\tau = t$ and work in the Newtonian limit. We should emphasize however that what follows does not rely in any crucial way on this limit and can be extended to include Post-Newtonian corrections, as we will discuss elsewhere~\cite{futureus}. The action (\ref{SO(3) action}) reduces to
\be \la{Newton action}
S &=& \int dt \bigg\{ \fr{mv^2}{2} - m \Phi + \fr{I}{2} \Omega_i \Omega^i + \fr{J}{4} (\Omega_i \Omega^i)^2 + \cdots \nonumber \\
&& \qquad + \fr{n_\Omega}{2} \Omega^i \Omega^j \d_i \d_j \Phi +  \fr{n_E}{4} \d^i \d^j \Phi \d_i \d_j \Phi + \cdots \\
&& \qquad \quad + \fr{1}{2} \d_i \d_j \Phi R^i{}_k R^j{}_l \tilde{\mathcal{O}}_E^{kl} + \cdots \bigg\}, \nonumber 
\ee
where $\Phi$ is the Newtonian potential and $\Omega^i = \fr{1}{2} \epsilon^{ijk}R_{jl} \d_t R_k{}^l$ is now the angular velocity in the inertial/lab frame. Each term in the action above lends itself to a simple physical interpretation. The first three terms amount to the usual Lagrangian for a non-relativistic spinning point particle coupled to gravity. The fourth term describes the deformation experienced by an object due to the centripetal force~\cite{Delacretaz:2014oxa}. In fact, expanding around a configuration with angular velocity $\bar \Omega^i$ yields an anisotropic correction to the inertia tensor $\delta I \propto J \bar \Omega^2 \simeq I (\bar \Omega / \Omega_0)^2$. Similarly, the first term in the second line describes the coupling between gravity and the ensuing quadrupole $\delta Q \propto \bar \Omega^2$. Alternatively, if we expand the gravitational potential around some background value $\bar \Phi$, we can also view this term as an additional deformation of the inertia tensor of the form $\delta I \propto n_\Omega \d^2 \bar \Phi$. The last term in the second line describes instead the coupling of the induced quadrupole $\delta Q \propto n_E \d^2 \bar \Phi$ to gravity.

Before varying the action (\ref{Newton action}) with respect to $x^i$ and $\theta^i$ to obtain the equations of motion, we should average over the degrees of freedom that we are not interested in explicitly keeping, which are encoded in the composite operator $\tilde{\mathcal{O}}_E^{ij}$. As was already recognized in the context of NRGR~\cite{Galley:2009px} and other dissipative systems~\cite{Endlich:2012vt}, such an averaging procedure can be implemented in a systematic way using the \emph{in-in formalism}~\cite{Jordan:1986ug}. In our context, this amounts to deriving the equations of motion by performing the following ``modified variation''~\cite{Galley:2012hx, futureus}:
\be \la{variation}
\delta S + i \int dt dt' \delta J_{ij}(t) \tilde G_R^{ijkl} (t - t') J_{kl}(t') = 0
\ee
where $ J_{kl} = \fr{1}{2} \d_i \d_j \Phi R^i{}_k R^j{}_l$ and $\tilde G_R$ is the retarded correlation function of the operators  $\tilde{\mathcal{O}}_E^{ij}$. If we assume that the degrees of freedom described by the $\tilde{\mathcal{O}}_E$'s are in near-equilibrium, then very general considerations imply that the Fourier transform $\tilde G_R (\omega)$ must be an odd, analytic function of $\omega$ that is positive for $\omega>0$~\cite{Endlich:2012vt}. Therefore, at low-frequencies we must have
\be \la{GR}
\tilde G_R^{ijkl} (\omega) \simeq \eta_E \omega (\delta^{ik} \delta^{jl} + \delta^{il} \delta^{kj} - \tfrac{2}{3} \delta^{ij} \delta^{kl}),
\ee
with $\eta_E \geqslant 0$. Since the $\tilde{\mathcal{O}}_E$ are quantities in the comoving frame, this correlation function is independent of spin, unlike the ones used in~\cite{Porto:2007qi}, which makes our theory more predictive. Moreover, $\eta_E$ can be extracted from numerical viscoelastic/hydrodynamic simulations for objects that are non-spinning. 

By combining equations (\ref{variation}) and (\ref{GR}) and varying w.r.t. $x^i$ and $\theta^i$ we finally obtain:
 \be
 && m \, \dot{v}_i = - m \d_i \Phi+ \fr{n_\Omega}{2} \Omega^k \Omega^j \d_i \d_j \d_k \Phi + \fr{n_E}{2} \d_i \d_j \d_k \Phi \d^j \d^k \Phi \nonumber \\
 && \qquad \quad    - \fr{\eta_E}{2}  \d_i \d_j \d_k \Phi ( \d^j \d^k \dot{ \Phi} + 2\d^j \d_l \Phi \epsilon^{klm} \Omega_m )\la{x eq}\\
&&  \d_t( I \Omega^i + J \Omega_j \Omega^j \Omega_i +n_\Omega \Omega^j \d_i \d_j \Phi ) =  -n_\Omega \epsilon_{ijk} \Omega^k \Omega^l \d^j \d_l \Phi \nonumber \\
&& + \eta_E \d^j \d^k \Phi (3 \d_i \d_j \Phi \Omega_k - 2 \d_j \d_k \Phi \Omega_i + \epsilon_{ikl} \d^l \d_j \dot{\Phi} ). \la{theta eq}
 \ee
The quantity in parentheses on the LHS of eq. (\ref{theta eq}) is simply the rotational angular momentum, which receives contributions from the terms proportional to $J$ and $n_\Omega$ as well. As we can see, this quantity is in general not conserved. However, it is easy to check that, in the case of a spherically symmetric potential $\Phi$, the orbital angular momentum also changes in such a way that the total angular momentum is conserved. The total energy is instead not conserved by an amount proportional to the dissipative coefficient $\eta_E$. In addition to the general framework, equations (\ref{x eq}) and (\ref{theta eq}) are the main results of this paper.

We pause here to note that while (\ref{x eq}) and (\ref{theta eq}) have terms which are quadratic in the Newtonian potential, and hence quadratic in Newton's constant $G$, they are not a post-Newtonian effect. That is, they survive even in the formal $c\rightarrow \infty$ limit and the elimination of any gravity self coupling.
\\


\noindent \textsc{Tidal Processes.} In a general scenario, the Newtonian potential $\Phi$ in equations (\ref{x eq}) and (\ref{theta eq}) depends on the coordinates of all other objects interacting gravitationally with the one under consideration, and they in turn obey similar equations of motion. In this letter, we will restrict ourselves to a simple setting in which a single object is moving in the gravitational field generated by another object with mass $M \gg m$, in which case we can neglect the motion of the latter. This amounts to setting $\Phi = - GM / r$ in (\ref{x eq}) and (\ref{theta eq}), which then reduce to
\be
&& m \, \dot{\vec{v}} = - \fr{GmM \hat{r}}{r^2} - \fr{9 n_E G^2 M^2  \hat r}{r^7} \nonumber \\
&& \qquad\quad  - \fr{3n_\Omega GM}{r^4} \l[\vec \Omega (\hat r \cdot \vec \Omega) + \frac{\hat r \Omega^2}{2} -\fr{5\hat r (\hat r \cdot \vec \Omega)^2 }{2}\r] \nonumber \\ 
&& \qquad\quad  - \fr{9\eta_E G^2 M^2}{r^8} \l[\vec r \times \vec \Omega + \vec v + 2 \hat r (\hat r \cdot \vec v) \r]\la{eq v central}  \\
&& \d_t \l\{ I \vec \Omega + J \Omega^2 \vec \Omega + \fr{GM n_\Omega}{r^3}[\vec \Omega - 3 \hat r(\hat r \cdot \vec \Omega) ] \r\} = \nonumber \\
&& \qquad \quad = \fr{3 n_\Omega G M (\hat r \cdot \vec \Omega) \hat r \times \vec \Omega}{r^3} \nonumber \\
&& \qquad \qquad + \fr{9\eta_E G^2 M^2}{r^7}[\hat r \times \vec v - r \vec \Omega + \hat r (\vec r \cdot \vec \Omega) ]. \la{eq omega central}
\ee
At this point, a few comments are in order. First, notice that these equations describe the \emph{instantaneous} change of velocity and angular momentum. Our derivation did not rely on any assumption about the direction of $\vec \Omega$ or about the shape of the orbit. In particular, we did not assume a closed orbit and did not average over a single period. Second, these equations are valid in the non-relativistic, slowly spinning limit and follow purely from symmetry considerations, supplemented with the assumption that the degrees of freedom that we are neglecting are in near-equilibrium. Last, our equations depend in total on six parameters (including the mass $m$ and moment of inertia $I$). Any detailed model for tidal dissipation corresponds to a specific choice for the coefficients $J, n_E, n_\Omega$ and $\eta_E$. For instance, the weak friction model~\cite{hut1981tidal} corresponds to setting $J = n_\Omega = 0$, $\eta_E = \tau n_E$ and $n_E = k \ell^5 /3 G$, where $\ell$ is the radius of the object, and $\tau$ and $k$ are respectively the time lag and apsidal motion parameters.

%
%

Equations (\ref{eq v central}) and (\ref{eq omega central}) become particularly simple when $\vec \Omega$ is perpendicular to the plane of the orbit:
\be
&& m \, \dot{\vec{v}} = - \fr{GM \hat{r}}{r^2} \l[ m+\fr{3 n_\Omega \Omega^2}{2 r^2} +\fr{9 n_E GM}{r^5} +\fr{27 \eta_E GM\dot r}{r^6}  \r] \nonumber \\
&& \qquad \qquad \qquad\qquad \qquad  -\fr{9 \eta_E G^2 M^2(\dot \theta - \Omega) \hat \theta}{r^7}, \la{eq v} \\
&& \d_t \l[ \l( I + J \Omega^2 + \fr{GM n_\Omega}{r^3}\r) \Omega \r] = \fr{9 \eta_E G^2 M^2 (\dot \theta - \Omega)}{r^6}. \la{eq Omega}
\ee
It is easy to see that a circular ($\dot r=0$) and tidally locked ($\dot \theta = \Omega$) orbit is a solution to these equations. We can also consider small perturbations around such a solution with radius $a_0$ and frequency $\Omega_0$. Taking the scalar product of (\ref{eq v}) with $\vec v$ gives an equation for $\dot E$ (where by $E$ we mean the sum of kinetic plus Newtonian potential energy), whereas taking the cross product with $\vec r$ yields an equation for the time derivative of the orbital angular momentum, $\dot{\vec{L}}$. Since for an elliptic orbit with semi-major axis $a$ and ellipticity $e$ we have $E = - GMm/2a$ and $L = \sqrt{GMm^2a(1-e^2)}$, we can easily derive equations for $\dot e$ and $\dot a$. Then, because the radius $r$ and the orbital angle $\theta$ are related by
\be
r = \fr{a (1-e^2)}{1+e \cos\theta},
\ee
we can average over one period to obtain 
\be
\dot{\delta e} \simeq -\fr{\delta e}{T_e},  \quad \fr{\dot{\delta a}}{a_0} = - \fr{2}{\alpha} \fr{\dot{\delta \Omega}}{\Omega_0} \simeq \fr{2}{7 T_e} \l( 3 \fr{\delta a}{a_0} + 2 \fr{\delta \Omega}{\Omega_0}\r), \qquad 
\ee
where the time scale of circularization $T_e$ and the parameter $\alpha$ are defined as follows:
\be
T_e = \fr{7 m a_0^8}{18 \eta_E G^2 M^2}, \qquad \alpha = \fr{m a_0^2}{I}.  
\ee
Using the fact that $\delta a$ and $\delta \Omega$ are related by conservation of the total angular momentum, one can also calculate the typical time scale of variation for $a$ and $\Omega$, $T_a$ and $T_\Omega$ respectively, and find
\be
T_a = T_\Omega = \fr{7 T_e}{2 \alpha} \propto T_e \, \fr{\ell^2}{a_0^2}
\ee
\\
Our effective theory approach makes it clear that this hierarchy of time scales is a model-independent result, which could indeed be confirmed by more precise observational data~(see for instance \cite{murray1999solar} pp.~166-173). Finally, for a tidally locked and slightly elliptical orbit one can easily integrate the equation for $\dot L$ to find how the orbital angular momentum varies along the orbit (being turned into spin angular momentum):
\be
L \simeq \sqrt{GMm^2a_0} - \fr{18 \eta_E G^2 M^2 e }{a_0^6} \, \sin \theta + \mathcal{O}(e^2). \quad 
\ee

To conclude, in this letter we have discussed a framework for studying tidal effects in astrophysics based on EFT techniques. While for simplicity we have considered spherically symmetric objects in the Newtonian limit, our formalism is very general and can systematically account for departures from sphericity as well as post-Newtonian and higher order finite-size corrections, as will be shown in~\cite{futureus}.
\\

\noindent \textsc{Appendix: the spin supplementary condition.} For readers well versed in the literature of relativistic spinning objects, the lack of a spin supplementary condition (SSC) will seem strikingly odd. In this short appendix, we comment on the apparent absence of such a condition in our formalism and attempt to make a connection with the standard approach to spinning objects.  

The dynamics of relativistic objects is usually described using the covariant 4-velocity $u^a$ and the (anti-symmetric) angular velocity tensor $\Omega^{ab}$, or equivalently, their conjugate momenta $p_a$ and $S_{ab}$ (see e.g. \cite{Porto:2007qi}). This covariant formulation introduces redundant degrees of freedom beyond the 3-velocity $v^i$ and Euler angles $\theta^i$ that are truly necessary to describe the position and orientation of the object. These additional degrees of freedom are of course not physical, as signaled by the fact that the covariant action is  invariant under some local (gauge) transformations. As usual, this gauge invariance can be dealt with by imposing suitable gauge-fixing conditions. This is exactly what happens in electromagnetism, where a Lorentz covariant description can be achieved only by introducing gauge invariance. More to the point (pun intended), in the case of a point particle without spin the gauge transformations in question are  reparametrizations of the world line. These are usually taken care of by imposing the ``gauge fixing condition'' $u_a u^a=-1$ to remove one spurious degree of freedom. 
 
In the case of spinning objects, one also needs to remove fictitious degrees of freedom from the relativistic angular momentum $S_{ab}$ or angular velocity $\Omega_{ab}$. The gauge-fixing condition one imposes in this case is know as \emph{spin supplementary condition} (SSC).\footnote{The formalism used in~\cite{Delacretaz:2014oxa} makes the gauge invariance associated with spinning objects particularly manifest, as argued in~\cite{Nicolis:2013sga}.} Just as the four velocity constraint $u_a u^a=-1$ has some physical meaning (the world line is parametrized by the rest frame time of the moving object), the same is true for different choices of the SSC. In fact, different SSC's correspond to different choices of the ``center of momentum'' about which the rotation is being defined. One standard choice for the SSC is $p_a S^{a b}=0$ where, again, $S_{a b}$ is the anti-symmetric angular momentum tensor and $p_a$ is the four momentum~\cite{Hanson:1974qy, Dixon, Pryce62}. From an EFT perspective, this choice is not very convenient. That is because both $S_{a b}$ and $p_a$ are defined as variations of the Lagrangian~\cite{Porto:2005ac} w.r.t. $\omega^{ab}$ and $u^a$ respectively. As such the condition $p_a S^{a b}=0$ depends on the form of the Lagrangian and thus must be amended anytime higher order corrections are added. From this viewpoint, imposing a condition on $\Omega^{ab}$ and $u^a$ seems a more natural choice. 

In our language, from equation~(\ref{omega_def}) we are furnished with the object $\tilde{\Omega}^{ij}=-\epsilon^{ijk}\tilde \Omega_k$. As it is defined in the rest frame of the object, it does not transform ideally under a rotation of the lab frame. However, we could consider instead $\Omega^{ij} \equiv \Lambda^i{}_k \Lambda^j{}_l \tilde{\Omega}^{ij}$, which transforms correctly under rotations. Based on the construction performed in \cite{Delacretaz:2014oxa}, this suggests that the natural Lorentz invariant object to consider is 
\begin{eqnarray}
\Omega^{cd}&\equiv& \Lambda^c{}_e \Lambda^d{}_f\tilde{\Omega}^{ef} \\
&\equiv& \Lambda^c{}_e \Lambda^d{}_f\left[\Lambda_a{}^e (\eta^{ab} \d_\tau + \d_\tau x^\mu \omega_\mu{}^{ab}) \Lambda_b{}^f\right]\\
&=&-\Lambda^c{}_a \d_\tau \Lambda^d{}^a + \d_\tau x^\mu \omega_\mu{}^{cd} \; \la{Omegacd},
\end{eqnarray}
where $\tilde \Omega^{i 0} =0$ by definition. We remind the reader that there is no need to impose additional constraints on $\Omega^{cd}$, because $\Lambda^a{}_b = B^a{}_c (\beta^i) R^c{}_b (\theta^i)$ where the $\theta$'s are Euler angles describing the orientation of the rigid body and $\beta^i$'s are fixed by equation~(\ref{beta_condition}). In other words, the covariant-looking quantity $\Omega^{cd}$ has been already gauge-fixed. We can reverse-engineer the gauge fixing condition by combining eq. (\ref{Omegacd}) with the fact that $u^a = \Lambda^a{}_0$~\cite{Delacretaz:2014oxa} to obtain
\be
\label{SSC_counterpart}
u_a \Omega^{ab}= - u^\mu \nabla_\mu u^a.
\ee
This equation can be thought of as the analog of the SSC in our formalism. Notice that the RHS of (\ref{SSC_counterpart}) has the geometrical interpretation of the extrinsic curvature of the world line~\cite{Delacretaz:2014oxa}. Equation (\ref{SSC_counterpart}) should be thought of as a geometrical constraint that holds independently of the equations of motion (it is valid ``off-shell'' in field theory parlance). Therefore, eq. (\ref{SSC_counterpart}) is not the same as $u_a \Omega^{ab}=0$ even though at lowest order, and in the absence of non-gravitational forces, the right hand side evaluates to zero on the equations of motion.
\\




\noindent \textsc{Acknowledgments.} We are extremely grateful to W.~Goldberger, L. Hui,  D.~Kipping, N.~Leigh, A.~Nicolis,  J.~Oishi, R.~Oppenheimer, I.~Rothstein and especially C.~Scharf and A.~Veicht for many enlightening conversations. We would also like to thank A.~Monin for his collaboration in the early stages of this project, and acknowledge the hospitality of the Sitka Sound Science Center where this work was first initiated.  This work was supported by the DOE under contract DE-FG02-11ER41743.

\bibliographystyle{apsrev4-1}
\bibliography{biblio}

\end{document}